\documentclass{PoS}
\usepackage{cite}
\usepackage{amsmath}
\def \picturesize {0.40\textwidth}

\title{Searches for long-lived and highly-ionizing particles at the CMS and ATLAS experiments}

\ShortTitle{Searches for long-lived and highly-ionizing particles}

\author{\speaker{Teresa Lenz} (on behalf of the ATLAS and CMS collaborations)\\ 
        Hamburg University\\
        E-mail: \email{teresa.lenz@desy.de}}

\abstract{
Long-lived particles are contained in a variety of beyond Standard Model theories, including supersymmetric models, universal extra dimensions, or technicolor theories. 
If the lifetime of such a particle is long enough, the particle can enter - or even pass through - the detector before it decays. 
Therefore, searches for long-lived particles require a very different search strategy compared to conventional searches for particles beyond the Standard Model. 
If the new particle is not only weakly interacting, the particle can be reconstructed itself and not only via its decay products. 
A very specific characteristic of such new heavy charged particles is their large ionization losses when traveling through the detector. 
This article summarizes searches for long-lived particles at the CMS and ATLAS experiments that exploit the potentially high ionization losses per path length ($dE/dx$) of the new particle. 
The presented searches are performed on 8 and/or 13\,TeV data. 
Additionally, an overview of the methodology of $dE/dx$ measurements at the CMS and ATLAS experiments is given.
}

\FullConference{Fourth Annual Large Hadron Collider Physics\\
		13-18 June 2016\\
		Lund, Sweden}

\begin{document}

\section{Introduction}
Long-lived particles are contained in many different extensions of the Standard Model (SM), including supersymmetric models, universal extra dimensions, or technicolor theories~\cite{bib:theory:SMP_at_colliders}. 
In these models the long lifetime is typically caused by suppressed couplings, by highly virtual processes, 
by a small phase space because of an approximate mass-degeneracy between two particles, or by (approximately) conserved quantum numbers.
If the lifetime is long enough, the corresponding particle can enter - or even pass through - the detector before it decays. 
Therefore, searches for long-lived particles require a very different search strategy compared to conventional searches for particles beyond the Standard Model (BSM).

A variety of different search strategies are exploited at the CMS~\cite{bib:CMS:detector} and ATLAS~\cite{bib:ATLAS:detector} experiments in order to search for neutral 
(e.g.~\cite{bib:CMS:Paper:DijetSearch,bib:ATLAS:Paper:DijetSearch,bib:ATLAS:Paper:LeptonJetsSearch,bib:CMS:PAS:DisplacedLeptons,bib:CMS:Paper:DisplacedLeptons,bib:ATLAS:Paper:DisplacedVertices,bib:ATLAS:Paper:DisplacedPhotons}) 
and charged (e.g.~\cite{bib:CMS:PAS:HSCP_13TeV,bib:ATLAS:Paper:Pixel+ToF_13TeV,bib:ATLAS:Paper:PixelDeDx_13TeV,bib:ATLAS:Paper:MultiChargedParticles_8TeV,bib:ATLAS:Paper:MagneticMonopoles_8TeV,bib:CMS:Paper:StoppedParticles,bib:ATLAS:Paper:StoppedParticles}) BSM particles.
These searches are designed in a way that they are able to cover a broad range of different lifetimes of the BSM particle leading to different decay points in the detector.

\begin{figure}[b]
 \centering
 \includegraphics[width=0.46\textwidth]{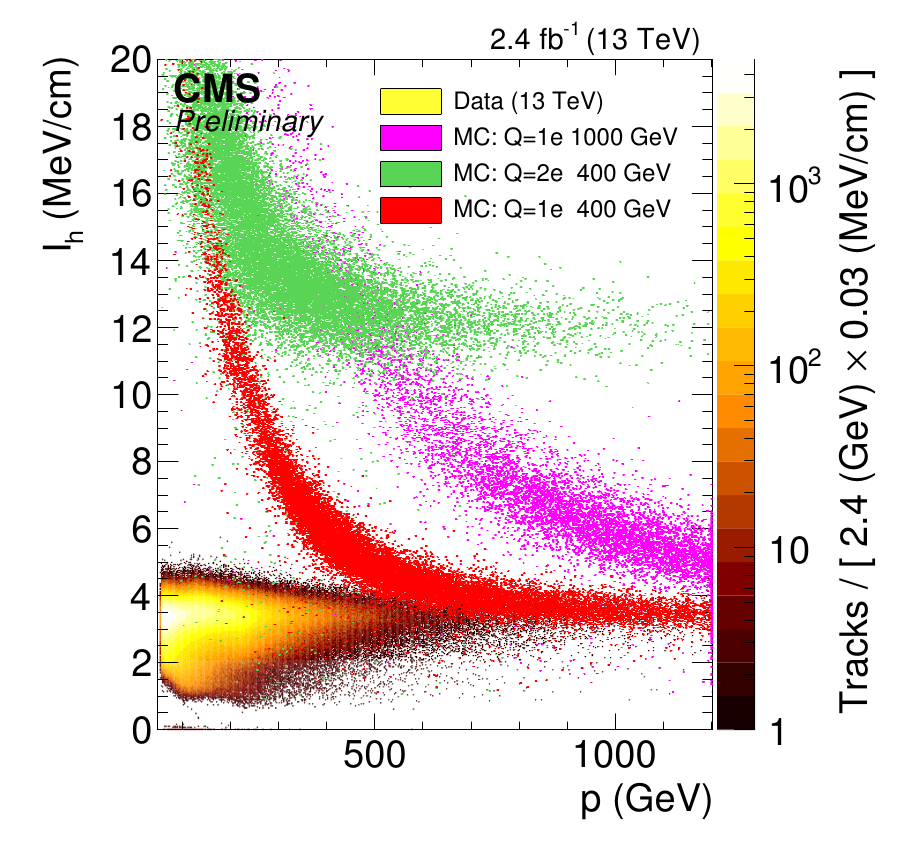}
 \caption{The most probable energy loss per path length $I_h$ (y-axis) vs. the momentum (x-axis) for data and three different signal models with single- and multiple-charged particles~\cite{bib:CMS:PAS:HSCP_13TeV}.}  
 \label{fig:dEdX_vs_P}
\end{figure}

If the new particle is not only weakly interacting, the particle can be reconstructed itself and not only via its decay products.
A very powerful discrimination against Standard Model particles can be achieved by exploiting the typically high energy loss per path length ($dE/dx$) of such a new heavy charged particle. 
The mean energy loss per path length $\langle dE/dx \rangle$ depends on various detector material parameters, and the velocity and charge of the incident particle. 
Thus, low velocities and multiple charges of particles lead to higher energy losses in the detector.
In contrast, high-momentum Standard Model particles which are light and therefore travel with very high velocities are only minimally ionizing.
In Fig.~\ref{fig:dEdX_vs_P} a comparison of Standard Model particles (data) and three different signal models is shown.
It can be seen, that the Standard Model background is distributed at $dE/dx$ values of around 3 MeV/cm for all momenta, whereas the BSM particles deposit much higher amounts of energy in the detector, especially for lower momenta ($p < 500\,$GeV).

Before presenting searches at the CMS and ATLAS experiment that make use of $dE/dx$, a short introduction into the methodology of measuring $dE/dx$ at the two experiments is given.

\section{dE/dx measurements at the ATLAS and CMS experiments}

The energy loss per path length is typically measured in the tracking system of the CMS and ATLAS experiments. 
Since the design of the tracking system of the CMS and ATLAS detectors differ, the respective methodology will be explained separately in the following two subsections.

\subsection{dE/dx measurements at the CMS experiment}
\label{sec:dEdx_CMS}
The tracker of the CMS detector is fully silicon based and is located inside a 3.8\,T solenoid. 
It consists of a silicon pixel detector as inner part and a silicon strip detector as outer part.
The tracker has a coverage in pseudorapidity up to $\eta = 2.5$ and extends radially between 4.4\,cm and 1.1\,m. 
The electric signals which are induced by particles passing through the silicon sensors are converted to a digital signal by the readout electronics.
In order to get a well calibrated energy signal the tracking system is continuously subject to an energy calibration.
When a particle passes through the full tracking system, it leaves typically around 17 $dE/dx$ measurements.
These single energy measurements need to be combined to one $dE/dx$ estimator that characterizes the energy loss of the particle.
At the CMS experiment, the so-called harmonic-2 estimator and a $dE/dx$ likelihood discriminator are used.
The harmonic-2 estimator is defined as follows 
\begin{equation}
I_{\text{h}}=\left( \frac{1}{N}\sum_{i=1}^{N}(\Delta E_i/\Delta x_i)^{-2} \right)^{-1/2}
\end{equation}
with $N$ being the number of energy measurements and $\Delta E/\Delta x$ being the single energy measurement per path length in one tracker module.
The harmonic-2 estimator suppresses extraordinary high energy losses by the power of $-2$. 
Since the underlying probability distribution of the single $\Delta E/\Delta x$ measurements is a Landau distribution which is a highly asymmetric function, the power of $-2$ is desirable in order not to bias the estimator by the long asymmetric right tail of the Landau distribution.
The harmonic-2 estimator therefore corresponds to the most probable energy loss per path length.
It can be further used to reconstruct the mass of the incident particle as will be seen later.

In addition to $I_{\text{h}}$, the so-called Asymmetric Smirnov-Cram\'{e}r-von Mises discriminator is used. 
It determines the compatibility of a set of $dE/dx$ measurements with the $dE/dx$ hypothesis distribution of minimally ionizing particles (MIPs).
Since it is a likelihood discriminator it is distributed between 0 and 1, where 0 corresponds to good compatibility with the MIP hypothesis and 1 corresponds to a bad compatibility with the MIP hypothesis.
This discriminator is especially suited to discriminate MIPs against highly-ionizing particles.

\subsection{dE/dx measurements at the ATLAS experiment}

The ATLAS tracking system is a mixture of a silicon tracker and a transition radiation tracker (TRT).
The silicon tracker is subdivided into a pixel tracker consisting of four layers since 13\,TeV data taking and a strip tracker.
Together they extend radially between $3.3\,$cm and 50\,cm.
Afterwards the TRT extends up to 1.1\,m.
Similar to the CMS tracker, the ATLAS tracker covers a pseudorapidity range up to $\eta=2.5$. 

The determination of the $dE/dx$ of a track is done separately in the silicon tracker and the TRT.
Since the silicon strip tracker records only a binary version of the energy release, only the silicon pixel tracker can be used to determine the energy loss per path length.
A particle leaves on average four $dE/dx$ measurements in the pixel tracker at 13\,TeV data taking and three $dE/dx$ measurements at 8\,TeV data taking.
These single measurements are then combined to either the mean or the most probable energy loss per path length.
Furthermore, the so-called $dE/dx$ significance is estimated which compares the observed $dE/dx$ of a track to the expected mean $dE/dx$ of a high-momentum muon
\begin{equation}
 S(\text{dE/dx}) = \frac{dE/dx_{\text{track}} - \langle dE/dx_{\mu} \rangle }{\sigma(dE/dx_{\mu})}.
 \label{eq:dEdxSignificance}
\end{equation}
In the TRT, a particle leaves typically around 32 straw hits when passing through.
These single hits can be combined to a $dE/dx$ significance estimator as defined in Eq.~\eqref{eq:dEdxSignificance}.
Furthermore, the fraction of hits that pass a certain predefined energy threshold can be calculated.
At the ATLAS experiment, the estimation of the deposited energy is done by measuring the time-over-threshold (ToT).
This time is correlated with the energy loss and can thus be used to determine the energy loss per path length.

\section{Searches for long-lived charged particles}
At the CMS experiment, a search for long-lived charged particles~\cite{bib:CMS:PAS:HSCP_13TeV} exploits the low velocity of a hypothetical new particle.
A low velocity of a particle has two major consequences: first, an unusually high energy loss, and second, a long time-of-flight (ToF). 
The latter one is measured in the muon system of the CMS detector.
Therefore the search uses two different strategies. 
The first search strategy includes only tracker variables (``tracker-only'' analysis) with the main focus on the extraordinary high energy loss of the BSM particle.
The second strategy combines the measurement of a highly-ionizing track in the tracking system with the time-of-flight measurement in the muon system (``tracker+ToF'' analysis).
In the following, the focus will be set on the ``tracker-only`` selection.

\begin{figure}[b]
 \centering
 \includegraphics[width=\picturesize]{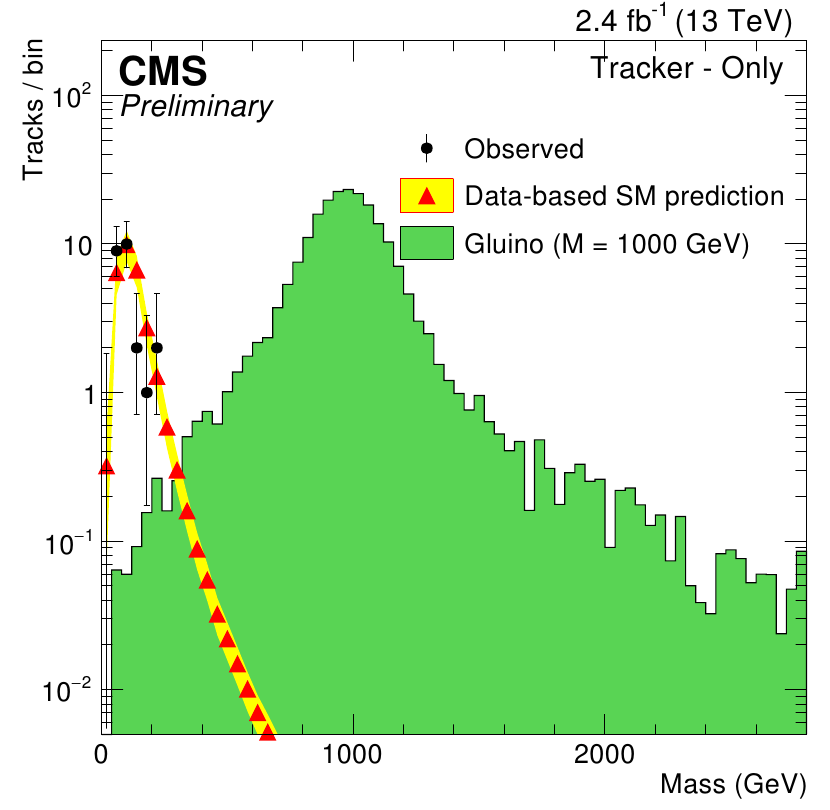}
 \hfill
  \includegraphics[width=\picturesize]{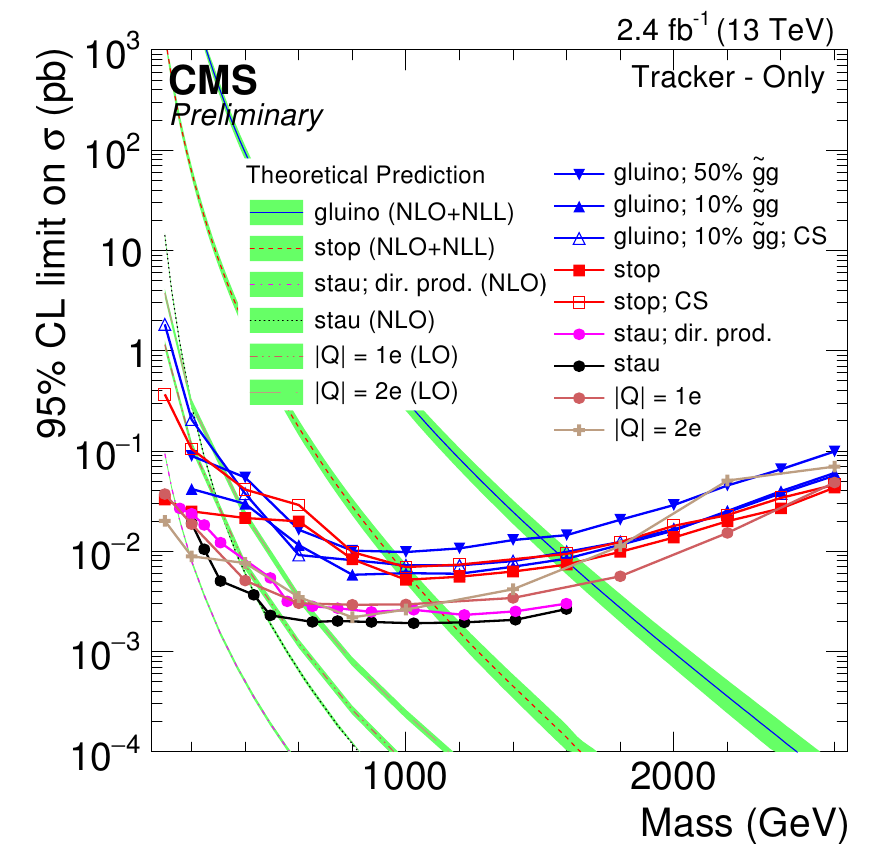}
 \caption{\textit{Left:} The predicted (red dots) and observed (black dots) mass spectrum of the ''tracker-only`` analysis of~\cite{bib:CMS:PAS:HSCP_13TeV}. A possible 1\,TeV gluino signal is indicated by the green area.
          \textit{Right:} The observed 95\% cross section upper limit for various signal models for the ''tracker-only`` analysis~\cite{bib:CMS:PAS:HSCP_13TeV}. The green lines indicate the theoretical cross sections.}  
 \label{fig:MassDistribution_HSCP_TrackerOnly}
\end{figure}

The ``tracker-only'' analysis makes use of a single-muon trigger and a missing energy ($E_{\text{T}}^{\text{miss}}$) trigger. 
The latter one is needed for particles traveling too slow to be reconstructed as muons, thus, inducing missing energy into the event.
In order to ensure fully efficient triggers, the missing energy is required to be larger than 170\,GeV and in case the particle is reconstructed as a muon, the muon transverse momentum to be larger than 50\,GeV.
In addition to this trigger selection, a candidate track selection is applied. 
This selection requires a good quality of the reconstructed track in order to reduce the background arising from fake tracks. 
Furthermore, isolation criteria in the tracking system and in the electromagnetic calorimeter are applied.
Finally, because of the typically high transverse momentum of a BSM particle, a $p_{\text{T}}$ of at least 55\,GeV is required.
The background to this search originates mainly from the tails of the $dE/dx$ distribution from MIPs. 
Since this happens very rarely, the background is very small and is not well described by simulated data. 
Therefore, this search relies on data-based background estimation methods.
The background is estimated with the help of control regions that are orthogonal to the signal region.
The final signal region is defined using the Asymmetric Smirnov-Cram\'{e}r-von Mises discriminator, $I_{\text{as}}$, (introduced in Section~\ref{sec:dEdx_CMS}) and the transverse momentum of the track, $p_{\text{T}}$.
The track is required to have \mbox{$I_{\text{as}}>0.3$} and \mbox{$p_{\text{T}}>65\,$GeV}.
The orthogonal regions are then defined with one (two) of these selection requirements inverted.
Since the variables are uncorrelated, the number of expected events in the signal region can be estimated with the number of events measured in the control regions in data $N_{\text{SR}} = \frac{ N_{\text{CR}_1}}{ N_{\text{CR}_2}} \cdot  N_{\text{CR}_3}$
where CR$_2$ corresponds to the control region with both requirements inverted and CR$_1$ and CR$_3$ are the control regions where only one of the selection requirements is inverted.
In addition, a mass reconstruction is done using an approximated expression of the Bethe-Bloch formula~\cite{bib:CMS:Paper:HSCP_7TeV}
\begin{equation}
\langle dE/dx \rangle = K \frac{m^2}{p^2} + C.
\end{equation}
The factors $K$ and $C$ are determined from low momentum protons in data.
Figure~\ref{fig:MassDistribution_HSCP_TrackerOnly} (\textit{left})  depicts the predicted and observed mass spectrum for the ''tracker-only`` analysis.
Since no excess in data is observed the search is interpreted in various SUSY models (split SUSY~\cite{bib:theory:SplitSUSY}, mGMSB~\cite{bib:theory:mGMSB}) for single-charged BSM particles. 
The search is also interpreted for double-charged particles which will be discussed in Section~\ref{sec:MultiChargedParticles}.
Figure~\ref{fig:MassDistribution_HSCP_TrackerOnly} (\textit{right}) shows the derived cross section upper limits for the ''tracker-only`` analysis.
The search excludes long-lived gluinos up to masses of 1590\,GeV, long-lived stops up to 1020\,GeV and long-lived staus up to 480\,GeV.
The limits could be increased with respect to the 8\,TeV analysis~\cite{bib:CMS:Paper:HSCP_8TeV} with a major improvement for long-lived gluinos.

\begin{figure}[!t]
 \centering
 \includegraphics[width=\picturesize]{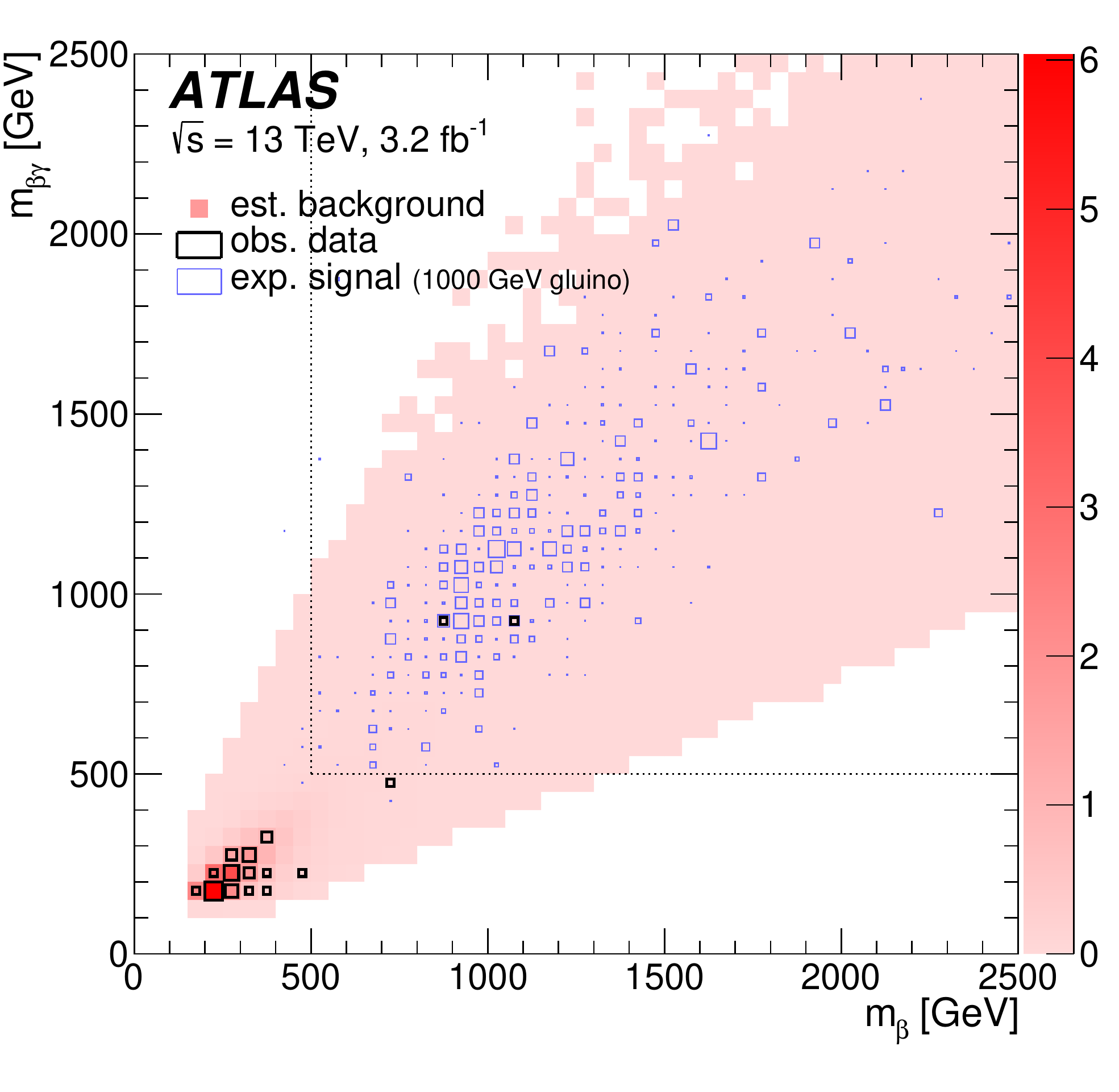}
 \hfill 
 \includegraphics[width=\picturesize]{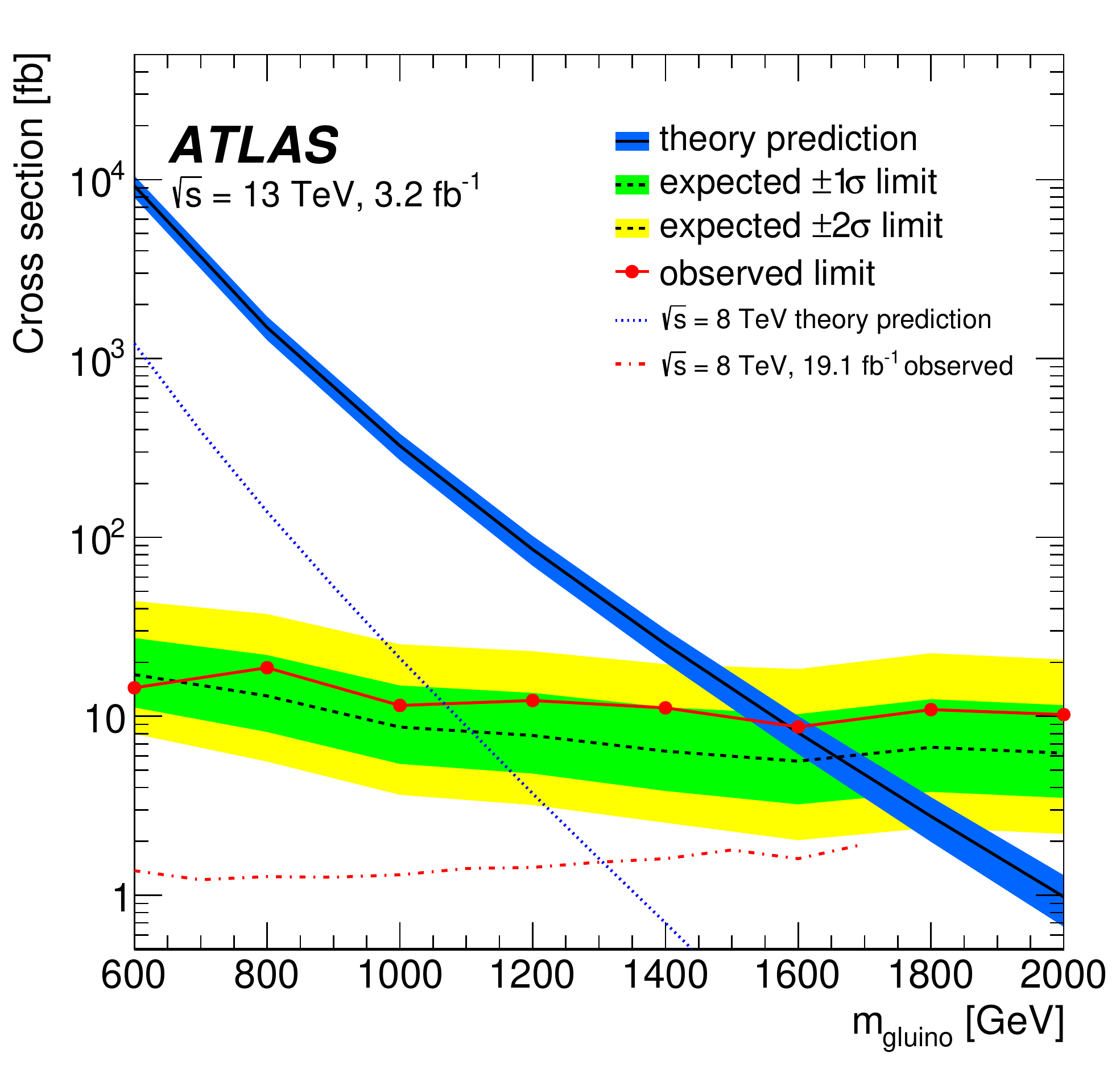}
 \caption{\textit{Left:} The reconstructed mass, $m_{\beta\gamma}$, derived from the $dE/dx$ measurements in the pixel tracker against the mass reconstructed from the time-of-flight measurement, $m_{\beta}$, for data, signal and expected background events~\cite{bib:ATLAS:Paper:Pixel+ToF_13TeV}.
                         The black dotted line indicates the signal region used for this specific signal model.
          \textit{Right:}~The observed (red line) and expected (black dashed line) 95\% cross section upper limit for stable gluinos~\cite{bib:ATLAS:Paper:Pixel+ToF_13TeV}. 
          The blue band indicates the theoretical cross section. 
          The green and yellow bands correspond to the 1 and 2\,$\sigma$ uncertainty of the expected upper limit, respectively.}  
 \label{fig:ATLAS:MassReconstruction}
\end{figure}

Similar searches at the ATLAS experiment also exploit ionization loss and time-of-flight measurements.
The search for heavy long-lived charged R-hadrons at 13\,TeV~\cite{bib:ATLAS:Paper:Pixel+ToF_13TeV} exploits alongside the measurement of $dE/dx$, the time-of-flight measured in the calorimeter. 
Both variables are then used to perform two independent mass reconstructions which in turn are used to discriminate a possible signal against the SM background.
Figure~\ref{fig:ATLAS:MassReconstruction} (\textit{left}) shows the reconstructed masses in the tracker and the calorimeter for data, the estimated background, and a 1\,TeV gluino signal.
The SM background of this search is estimated by sampling points from probability distribution functions, determined in control regions in data, in the variables $\beta$, $\beta\gamma$, and in the momentum $p$.
This method can be used because $dE/dx$, the time-of-flight and the momentum are uncorrelated for high-momentum SM particles.
The normalization of the estimated background is done outside the signal region, i.e. in the region where at least one of the mass requirements - imposed on the reconstructed masses using $dE/dx$ and ToF - fails.
The search is interpreted for various types of \mbox{R-hadrons} (gluinos, stops, sbottoms). 
Especially for models with long-lived gluinos the cross section upper limit has been significantly improved with respect to the 8\,TeV search~\cite{bib:ATLAS:Paper:Pixel+ToF_8TeV}.
Figure~\ref{fig:ATLAS:MassReconstruction} (\textit{right}) depicts the cross section upper limit for stable gluinos.

\begin{figure}[!b]
 \centering
 \begin{minipage}{0.49\textwidth}
 \includegraphics[width=0.99\textwidth]{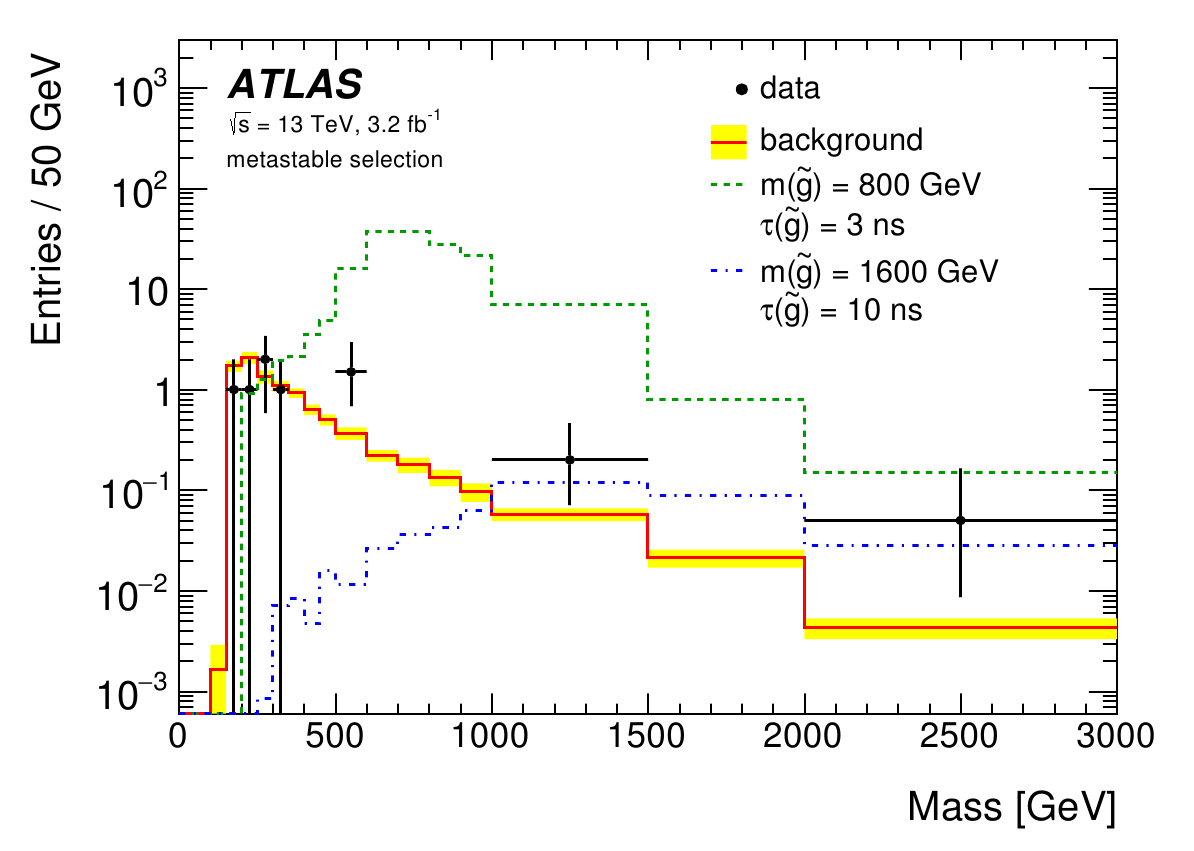}
 \end{minipage}
 \hfill
 \begin{minipage}{\picturesize}
 \includegraphics[width=0.99\textwidth]{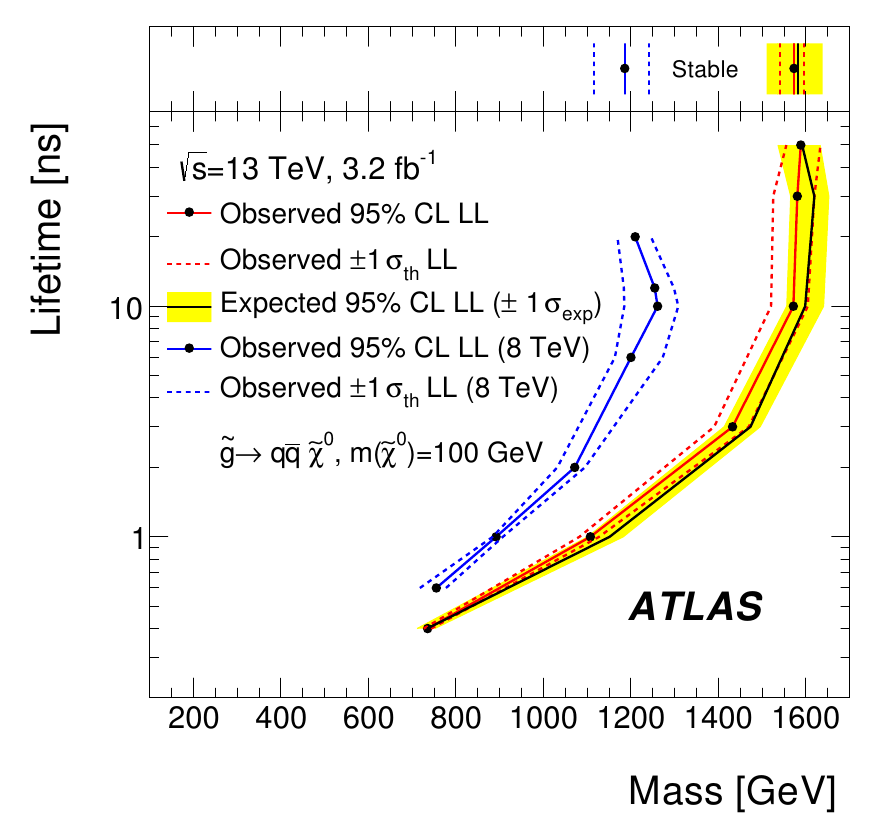}
 \end{minipage}
 \caption{\textit{Left:} The mass distribution of reconstructed tracks in the pixel tracker for background (red line), signal (dashed lines), and data (black dots) events~\cite{bib:ATLAS:Paper:PixelDeDx_13TeV}.
          \textit{Right:} The 95\% cross section upper limit for long-lived gluinos ($\tau>0.4\,$ns) decaying to $q\overline{q}$ and a 100\,GeV neutralino~\cite{bib:ATLAS:Paper:PixelDeDx_13TeV}.}  
 \label{fig:ATLAS:PixelDeDx}
\end{figure}

A second search at the ATLAS experiment concentrates on the search for heavy charged particles using the pixel $dE/dx$ measurement only~\cite{bib:ATLAS:Paper:PixelDeDx_13TeV}.
Again, a mass reconstruction is done with the help of the $dE/dx$ measurement in the pixel tracker.
Figure~\ref{fig:ATLAS:PixelDeDx} (\textit{left}) depicts the reconstructed mass for the background, a potential signal, and for data.
The background is estimated in a data-based approach by sampling from distributions of the key variables $p$, $dE/dx$, $\eta$ determined in control regions in data.
The normalization is done by scaling the sampled points to the number of events in the control region with $M<160\,$GeV.
This search is interpreted in mini-split supersymmetric models and AMSB models and excludes gluinos with a lifetime greater than 0.4\,ns up to 1590\,GeV (see Fig.~\ref{fig:ATLAS:PixelDeDx} (\textit{right})).

\section{Searches for multi-charged particles}
\label{sec:MultiChargedParticles}
\begin{figure}[b]
 \centering
 \begin{minipage}{0.49\textwidth}
 \includegraphics[width=0.99\textwidth]{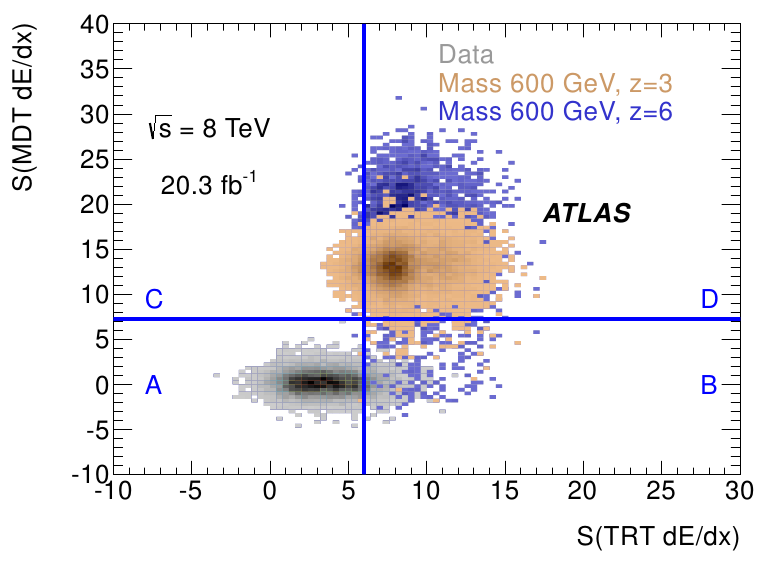}
 \end{minipage}
 \hfill
 \begin{minipage}{0.49\textwidth}
 \includegraphics[width=0.99\textwidth]{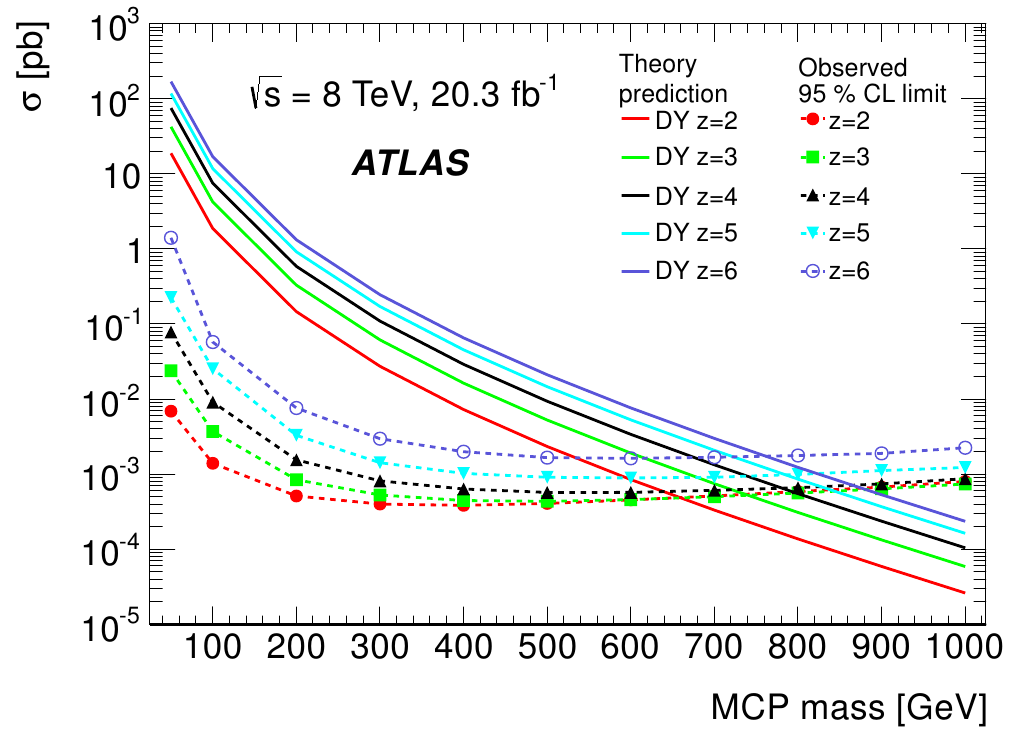}
 \end{minipage}
 \caption{\textit{Left:} The 2-dimensional distribution of the $dE/dx$ significance measured in the MDT (y-axis) and TRT (x-axis) for data and two different signal models~\cite{bib:ATLAS:Paper:MultiChargedParticles_8TeV}.
          \textit{Right:} The 95\% cross section upper limit for multi-charged particles (dashed lines) and their corresponding theoretical cross sections (solid lines)~\cite{bib:ATLAS:Paper:MultiChargedParticles_8TeV}.}  
 \label{fig:ATLAS:MultiCharged_ABCD_Limits}
\end{figure}

Furthermore, searches for multi-charged particles are performed.
The search for heavy long-lived multi-charged particles using the ATLAS detector~\cite{bib:ATLAS:Paper:MultiChargedParticles_8TeV} analyzes data taken at $\sqrt{s} =$8\,TeV. 
It makes use of four different $dE/dx$ estimators: $dE/dx$ significances measured in the pixel detector ($S_{\text{pixel}}$), in the TRT ($S_{\text{TRT}}$), and in the monitored drift tube chambers ($S_{\text{MDT}}$), 
as well as the fraction of hits passing a high energy threshold ($f^{\text{HT}}_{\text{TRT}}$) in the TRT.
Since this search concentrates on very long-lived particles, the initial selection is seeded by a reconstructed track in the muon system with at least 8 MDT hits.
Additionally, a candidate track selection is applied for reconstructed tracks in the inner tracking system. 
It requires an isolated high-quality track with at least $p_{\text{T}}>30$\,GeV for double-charged particles and $p_{\text{T}}>40$\,GeV for particles with $z \geq 3$.
The final selection is based on the $dE/dx$ significance measured in the pixel detector and $f^{\text{HT}}_{\text{TRT}}$.
The background is estimated from data in certain control regions using the uncorrelated variables $S_{\text{TRT}}$ and $S_{\text{MDT}}$.
The 2-dimensional distribution of these two variables for data and two different signal samples is shown in Fig.~\ref{fig:ATLAS:MultiCharged_ABCD_Limits} (\textit{left}).
The control regions indicated by the blue lines in Fig.~\ref{fig:ATLAS:MultiCharged_ABCD_Limits} (\textit{left}) are used to predict the number of background events in the signal region (region D).
This search predicts $0.013\pm0.002$ and $0.026\pm0.003$ events for the $z=2$ and $z \geq 3$ selections, respectively.
Since the number of observed events is zero in either of the two signal regions, thus,  compatible with the background prediction, cross section upper limits are set for Drell-Yan models with pure electromagnetic couplings.
The limits are shown in Fig.~\ref{fig:ATLAS:MultiCharged_ABCD_Limits} (\textit{right}) for particles with $z= 2-6$.

At CMS, a search for multi-charged particles is embedded in the search for long-lived charged particles~\cite{bib:CMS:PAS:HSCP_13TeV} presented before.
Double-charged particles are excluded up to masses of 650\,GeV. 
This limit is a bit lower compared to the 8\,TeV analysis~\cite{bib:CMS:Paper:HSCP_8TeV} which could excluded double-charged particles up to masses of 730\,GeV.
The difference is caused by the specific selection optimized for multi-charged particles used for the 8\,TeV analysis.

\section{Search for magnetic monopoles}
\label{sec:MagneticMonopoles}
\begin{figure}[b]
 \centering
 \begin{minipage}{0.49\textwidth}
 \includegraphics[width=0.99\textwidth]{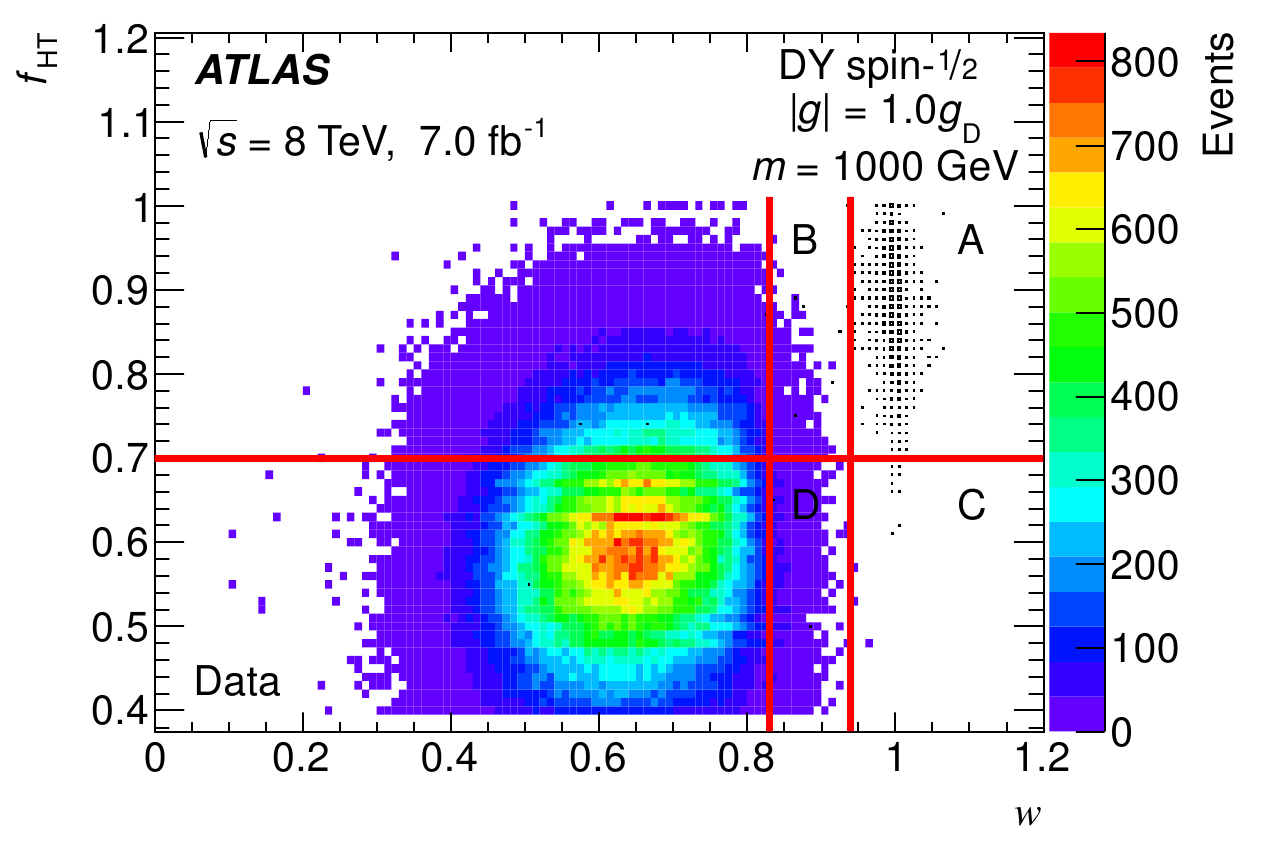}
 \end{minipage}
 \hfill
 \begin{minipage}{\picturesize}
 \includegraphics[width=0.99\textwidth]{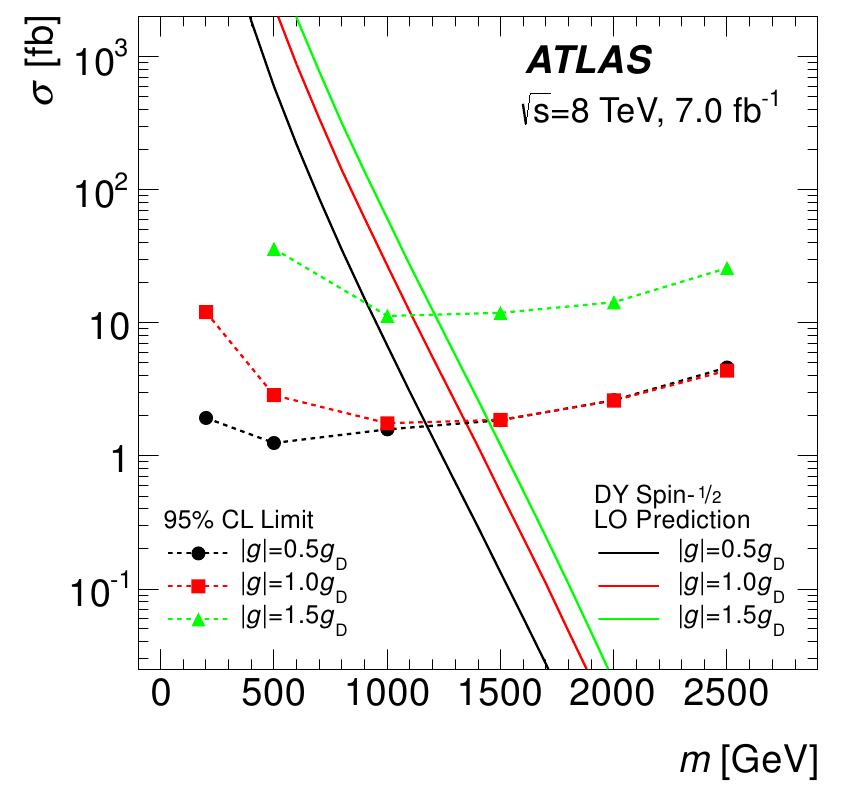}
 \end{minipage} 
 \caption{\textit{Left:} The 2-dimensional distribution of $w$ (x-axis) and $f^{\text{HT}}_{\text{TRT}}$ (y-axis) for data and a signal model with a mass of the magnetic monopole of $1\,$TeV and a spin of $1/2$~\cite{bib:ATLAS:Paper:MagneticMonopoles_8TeV}. 
                The definition of the signal region (A) and the control regions (B, C, D) is indicated by the red lines.
          \textit{Right:} The 95\% cross section upper limits (dashed lines) and the theoretical cross sections (solid lines) for magnetic monopoles with spin 1/2 and different magnetic charges~\cite{bib:ATLAS:Paper:MagneticMonopoles_8TeV}.}  
 \label{fig:ATLAS:Monopoles_f_w}
\end{figure}

The Dirac argument~\cite{bib:ATLAS:Paper:Dirac} offers a possible explanation for the quantization of the electric charge.
For this argument however, the existence of magnetic monopoles is necessary.
The magnetic charge, $g_D$, of monopoles can be related to the electric charge, $e$, with the following formula
\begin{equation}
 \frac{g_D}{e} = \frac{1}{2 \alpha_e} \approx 68.5.
\end{equation}
Thus, magnetic monopoles are characterized by huge ionization losses.
At the ATLAS experiment a search for magnetic monopoles is performed on 8\,TeV data~\cite{bib:ATLAS:Paper:MagneticMonopoles_8TeV}.
This search exploits a dedicated software trigger for highly-ionizing particles which makes use of two different $dE/dx$ variables: 
$f^{\text{HT}}_{\text{TRT}}$ (defined in the previous Section~\ref{sec:MultiChargedParticles}), and the number of hits in the TRT that pass a high energy threshold, $N^{\text{HT}}_{\text{TRT}}$.
The discriminating particle characteristics used by this search are the energy, the energy dispersion in the electromagnetic calorimeter, $w$, and $f^{\text{HT}}_{\text{TRT}}$. 
The energy dispersion measures the fraction of the cluster energy contained in the most energetic cells of a cluster in each of the layers of the electromagnetic calorimeters.
The two variables $w$ and  $f^{\text{HT}}_{\text{TRT}}$ are furthermore used to estimate the background in a data-based way.
Figure~\ref{fig:ATLAS:Monopoles_f_w} (\textit{left}) shows the 2-dimensional distribution of $w$ and $f^{\text{HT}}_{\text{TRT}}$ for data and a signal with $1.0\,g_D$.
It can be seen that the variables are well discriminating and can thus be used to predict the number of background events in the signal region (region~A).
Since $w$ and $f^{\text{HT}}_{\text{TRT}}$ are not fully uncorrelated for smaller values of $w$, the control regions B and D are defined with a lower bound of $w$ of around~0.8.
Since no excess of data events is observed, the search is interpreted in Drell-Yan production models with modified electromagnetic couplings (see Fig.~\ref{fig:ATLAS:Monopoles_f_w} (\textit{right})).
The search excludes magnetic monopoles with a magnetic charge of $1.0\,g_D$ up to masses of 1340\,GeV for a spin 1/2 hypotheses of the particle. 
In addition, a model independent cross section upper limit is determined in fiducial regions where the selection efficiency is almost constant.
This cross section upper limit is estimated to a value of 0.5\,fb.

\section{Conclusions}
Long-lived particle searches are sensitive to a variety of different models beyond the Standard Model. 
These searches have in common that they typically feature very low backgrounds.
Furthermore, because of the unconventional signatures, they need to rely on data-based background estimation techniques and require a good understanding of the detector.

So far, searches for long-lived particles at the LHC show no evidence for new physics. 
However, many scenarios haven't been accessible so far and will need to be tested with future LHC data.
Long-lived particle searches can thereby complement conventional searches and will therefore stay very important at the~LHC.

\section*{ACKNOWLEDGMENTS}
I thank the ATLAS and CMS collaborations for the work presented here and for the support preparing this talk, the CERN accelerator division for the excellent operation of the LHC, and all funding agencies that made these experiments possible.


\end{document}